\documentclass[manuscript,screen,nonacm]{acmart}
\usepackage{subfig}

\AtBeginDocument{%
  \providecommand\BibTeX{{%
    \normalfont B\kern-0.5em{\scshape i\kern-0.25em b}\kern-0.8em\TeX}}}

\setcopyright{acmcopyright}
\copyrightyear{2022}
\acmYear{2022}
\acmDOI{XXXXXXX.XXXXXXX}

\acmConference[AVI '22]{International Conference on Advanced Visual Interfaces, Workshop on Visuo-Haptic Interaction}{June 07,
  2022}{Rome, Italy}
\acmBooktitle{AVI '22: International Conference on Advanced Visual Interfaces, Workshop on Visuo-Haptic Interaction,
 June 07, 2022, Rome, Italy} 
\acmPrice{15.00}
\acmISBN{978-1-4503-XXXX-X/18/06}

\begin{document}

\title{Extending Cobot's Motion Intention Visualization by Haptic Feedback}

\author{Max Pascher}
\orcid{0000-0002-6847-0696}
\email{max.pascher@w-hs.de}
\affiliation{
    \institution{Westphalian University of Applied Sciences}
    \city{Gelsenkirchen}
    \country{Germany}
}
\affiliation{
    \institution{University of Duisburg-Essen}
    \city{Essen}
    \country{Germany}
}

\author{Til Franzen}
\orcid{0000-0003-0203-7512}
\email{til.franzen@studmail.w-hs.de}
\affiliation{
    \institution{Westphalian University of Applied Sciences}
    \city{Gelsenkirchen}
    \country{Germany}
}

\author{Kirill Kronhardt}
\orcid{0000-0002-0460-3787}
\email{kirill.kronhardt@studmail.w-hs.de}
\affiliation{
    \institution{Westphalian University of Applied Sciences}
    \city{Gelsenkirchen}
    \country{Germany}
}

\author{Jens Gerken}
\orcid{0000-0002-0634-3931}
\email{jens.gerken@w-hs.de}
\affiliation{
    \institution{Westphalian University of Applied Sciences}
    \city{Gelsenkirchen}
    \country{Germany}
}

\renewcommand{\shortauthors}{Pascher et al.}

\begin{abstract}
Nowadays, robots are found in a growing number of areas where they collaborate closely with humans. Enabled by lightweight materials and safety sensors, these cobots are gaining increasing popularity in domestic care, supporting people with physical impairments in their everyday lives.
However, when cobots perform actions autonomously, it remains challenging for human collaborators to understand and predict their behavior, which is crucial for achieving trust and user acceptance.
One significant aspect of predicting cobot behavior is understanding their motion intention and comprehending how they "think" about their actions. Moreover, other information sources often occupy human visual and audio modalities, rendering them frequently unsuitable for transmitting such information. 
We work on a solution that communicates cobot intention via haptic feedback to tackle this challenge. In our concept, we map planned motions of the cobot to different haptic patterns to extend the visual intention feedback.  
\end{abstract}



\begin{teaserfigure}
  \includegraphics[width=\textwidth]{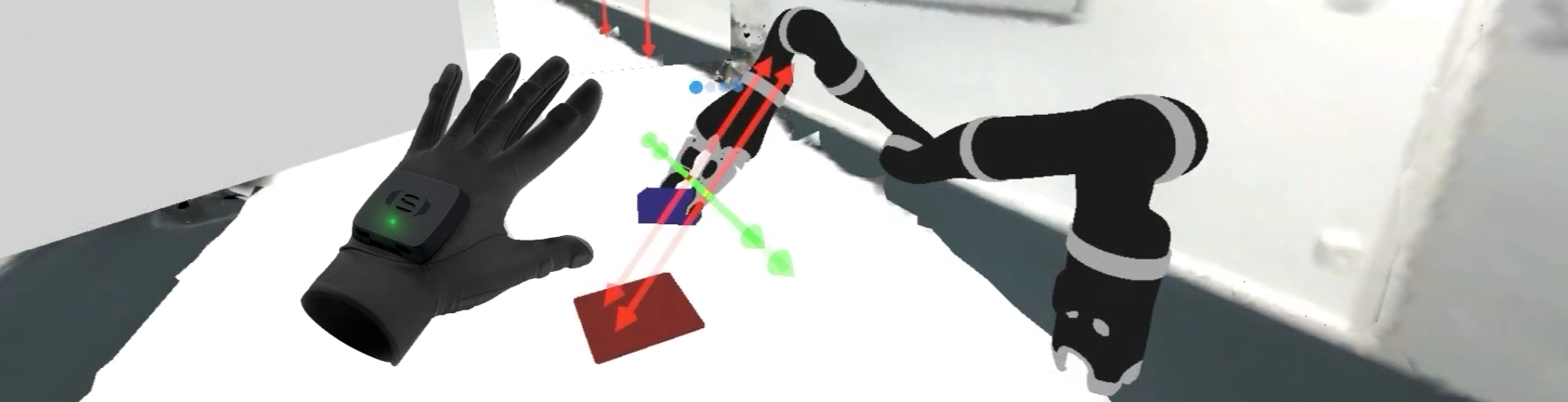}
  \caption{Improving human-robot collaboration by visual and haptic intention feedback.}
  \Description{In a virtual reality environment, a simulated robotic arm (cobot) gripped a blue box and wanted to place it on a red spot on a table. Green and red arrows provide a visual and a glove haptic intention feedback.}
  \label{fig:teaser}
\end{teaserfigure}


\maketitle

\section{Introduction}
Robotic solutions are becoming increasingly prevalent in our personal and professional lives, and have started to evolve into close collaborators~\cite{Bauer.2008, Fong.2003, ArevaloArboleda.2020}.
These so-called cobots support humans in various ways that were unimaginable just a few years ago. Enabled by technological advances, newer lightweight materials, and improved safety sensors, they are gaining increasing popularity in domestic care, supporting people with disabilities in their everyday lives.

A non-negligible number of people live with motor impairments, ranging from slight limitations to severe paralysis~\cite{Pflegestatistik}. While a near-complete integration into personal and social life is the final goal, current cobots focus on performing activities of daily living~\cite{Pascher.2019}. These include essentials like eating and drinking or more complex tasks such as grooming and activities associated with leisure time~\cite{Pascher.2021recommendations}.

However, new potential issues arise when cobots are tasked with autonomous or semi-autonomous actions, resulting in added stress for end-users~\cite{Pollak.2020}.
Particularly close proximity collaboration between humans and cobots remains challenging~\cite{gruenefeld2020mind}.
These challenges include effective communication to the end-user of (a) motion intent and (b) the spatial perception of the cobot's vicinity~\cite{Chadalavada.2015,Pascher.2022}. Accurate communication increases user understanding while avoiding the unpredictability regarding impending steps, motions, and sensed environment parameters.
While visualizations of motion intent have been extensively studied~\cite{Andersen.26.08.201631.08.2016,Chadalavada.2015,Coovert.2014,Stulp.28.09.201502.10.2015,Watanabe.28.09.201502.10.2015,gruenefeld2020mind}, communicating cobot intention via haptic has received less attention~\cite{Grushko.2021tactile,Grushko.2021haptic}. 

We investigate a new visual-haptic approach that communicates the cobot's intention, focusing primarily on information about its planned path, to the human collaborator (see~\autoref{fig:teaser}). Information about the path is crucial as, particularly in a close proximity collaboration situation, any misunderstanding in the cobot's motion intention can result in errors in behavior. These range from knocking over objects or even destroying them in the process to potentially harming the user. Minimizing these risks is an important step in the development of effective robotic solutions with wide end-user acceptance.

The complex-looking supporting actions of the cobot are often just a series of pick \& place tasks. In our test scenario the user sits in front of a table with an object on it and a cobot mounted to the surface. The cobot assists the user by picking up the object and placing it in a dedicated spot on the table.

\section{Related Work}
Previous literature has focused either on (a) visualization or (b) haptic techniques to communicate the cobot's motion intention to the user. Combining these two approaches we focus on ways cobots can effectively communicate their planned path with a visual-haptic solution. 

\subsection{Visualization Techniques to Communicate Cobot's Motion Intention}
In recent decades, Augmented Reality (AR) technology has been frequently used for human-robot collaboration~\cite{Dianatfar.2021}. Previous work focused mainly on the use of Head-Mounded Displays (HMDs), Mobile Augmented Reality (MAR), and Spatial Augmented Reality (SAR) for the visualization of the cobot motion intent~\cite{Rosen.2019,Walker.2018,gruenefeld2020mind}. 
Rosen et\,al. showed that AR is an improvement compared to classical desktop interfaces when visualizing the intended motion of robots~\cite{Rosen.2019}.
However, while visualizations of motion intent have been studied extensively in previous work~\cite{Andersen.26.08.201631.08.2016,Chadalavada.2015,Coovert.2014,Stulp.28.09.201502.10.2015,Watanabe.28.09.201502.10.2015,gruenefeld2020mind}, communicating cobot intention via haptic has not attracted as much attention.

\subsubsection{Techniques Involving Haptic to Communicate Cobot's Motion Intention}
Previous research used haptic and tactile feedback to guide users in a specific direction for example by providing vibration feedback~\cite{Lehtinen.2012,Barralon.2009,Chen.2018,Hong.2017,Weber.2011}.
Grushko et\,al. transferred these findings into the domain of human-robot collaboration by communicating directional information through the activation of six actuators on a glove spatially organized to represent an orthogonal coordinate frame~\cite{Grushko.2021tactile}. The vibration activates on the side of the glove that is closest to the future path of the robot.
They also use this haptic device to notify the user about the currently planned robot's trajectory and status changes~\cite{Grushko.2021haptic}.

\section{Approach}
\begin{figure}[htbp]
    \centering
    \subfloat[]{\includegraphics[width=0.7\linewidth]{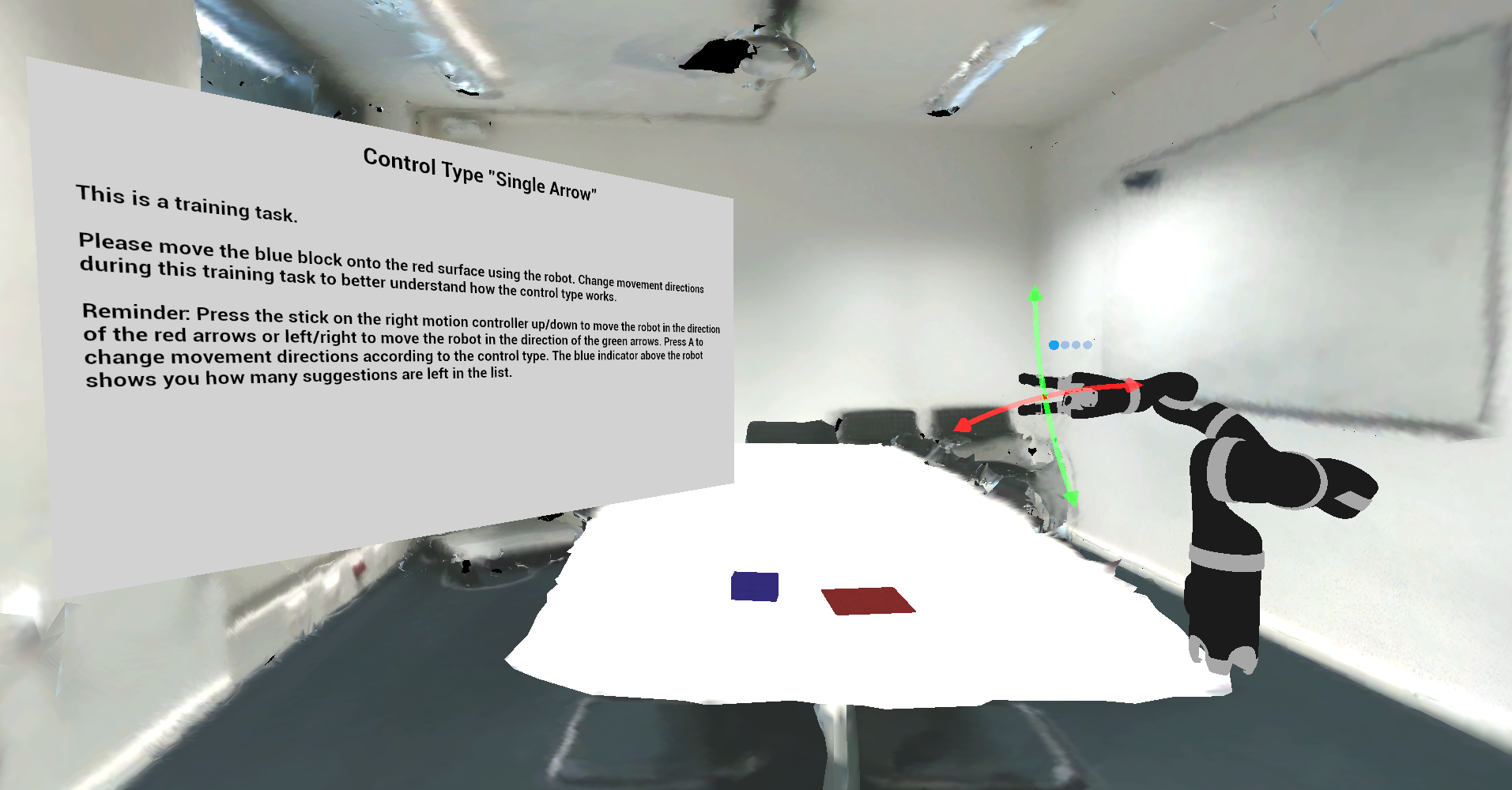}\label{fig:apparatus}}
    \hfill
    \subfloat[]{\includegraphics[width=0.268\linewidth]{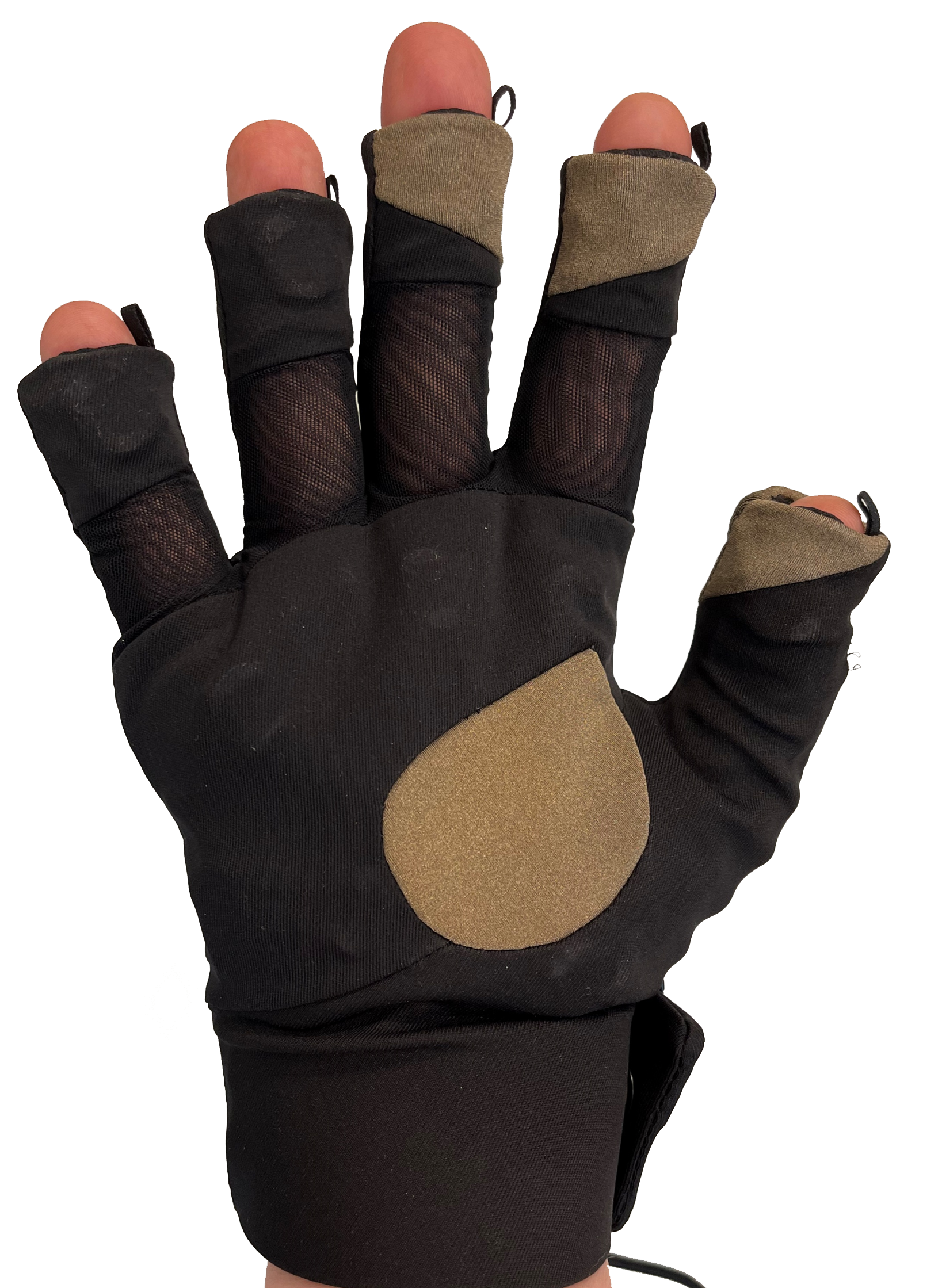}\label{fig:glove}}
    \caption{(a) The virtual environment: description screen (\textbf{Left}); \emph{Kinova Jaco} with visualisation for control type \emph{Single Arrow} (\textbf{Right}); table with blue block and red target (\textbf{Bottom}); (b) \emph{Sensorial XR} haptic glove.}
    \label{fig:overview}
\end{figure}

In earlier work, we developed an adaptive control interaction method based on a recommendation system generated by a Convolutional Neural Network (CNN)~\cite{Kronhardt.2022}. From the cobot's seven Degrees of Freedom (DoF), the adaptive control combined several DoFs to provide a more straightforward control to the user with fewer necessary mode-switches. We compared the novel adaptive control method, with two different visualization techniques, to the standard mode-switch approach with cardinal DoF mappings.
We also designed a Virtual Reality (VR) environment based on a photogrammetry scan of a physical room. The environment included a virtual model of the \emph{Kinova Jaco}\footnote{Kinova Jaco robot arm: \url{https://assistive.kinovarobotics.com/product/jaco-robotic-arm}, last retrieved \today} robot arm attached to a table, a red target area, and a blue block (see Figure~\ref{fig:apparatus}).

The virtual environment was developed to be compatible with the \emph{Oculus Quest 2}\footnote{Oculus Quest 2: \url{https://www.oculus.com/quest-2/}, last retrieved \today} VR headset. This provided us with a VR testbed environment for developing and evaluating further feedback techniques.
Currently we aim to develop multi-modal feedback methods for human-robot collaborations beyond visual and audio by providing haptic feedback. We are working on different concepts to communicate the cobot's motion intent via vibrotactile feedback by using the \emph{Sensorial XR}\footnote{Sensorial XR: \url{https://sensorialxr.com/}, last retrieved \today} (see Figure~\ref{fig:glove}).
To communicates directional information to the human collaborator different mappings of vibrotactile actuators correspond to matching DoF combinations of the adaptive control.
This brings the cartesian coordinate systems of the cobot and the glove in line to provide an intuitive mapping. Changes in the intensity of the actuators indicate the amount of directional change, thus enabling the user to better imagine the path generated by the recommendation system, resulting in low end-user task load.

\bibliographystyle{ACM-Reference-Format}
\bibliography{ms}

\end{document}